\documentclass[pdflatex,sn-mathphys-num]{sn-jnl}

\usepackage{graphicx}%
\usepackage{multirow}%
\usepackage{amsmath,amssymb,amsfonts}%
\usepackage{amsthm}%
\usepackage{mathrsfs}%
\usepackage[title]{appendix}%
\usepackage{xcolor}%
\usepackage{textcomp}%
\usepackage{manyfoot}%
\usepackage{booktabs}%
\usepackage{algorithm}%
\usepackage{algorithmicx}%
\usepackage{algpseudocode}%
\usepackage{listings}%
\usepackage{url}
\usepackage{hyperref}
\usepackage{fancyhdr}
\usepackage[absolute,overlay]{textpos}
\usepackage{tabularx}
    \newcolumntype{L}{>{\raggedright\arraybackslash}X}
\usepackage{tikz}
\usetikzlibrary{arrows.meta,positioning,fit,shapes.multipart}
\theoremstyle{thmstyleone}%
\theoremstyle{thmstyletwo}%

\theoremstyle{thmstylethree}%

\begin{document}
\begin{textblock}{16}(2,2)  
    \noindent Theodore S. Rappaport, Todd E. Humphreys, and Shuai Nie, ``Spectrum Opportunities for the Wireless Future: From Direct-to-Device Satellite Applications to 6G Cellular,'' (Invited) \textit{npj Wireless Technology}, 2025.
\end{textblock}

\title[Article Title]{Spectrum Opportunities for the Wireless Future: From Direct-to-Device Satellite Applications to 6G Cellular}

\author[1]{\fnm{Theodore S.} \sur{Rappaport}}\email{tsr@nyu.edu}

\author[2]{\fnm{Todd E.} \sur{Humphreys}}\email{todd.humphreys@utexas.edu}

\author*[3]{\fnm{Shuai} \sur{Nie}}\email{shuainie@unl.edu}

\affil[1]{\orgdiv{NYU WIRELESS}, \orgname{New York University}, \orgaddress{\city{Brooklyn}, \state{NY} \postcode{11201}, \country{USA}}}

\affil[2]{\orgdiv{Department of Aerospace Engineering and Engineering Mechanics}, \orgname{University of Texas at Austin}, \orgaddress{\city{Austin},  \state{TX} \postcode{78712}, \country{USA}}}

\affil*[3]{\orgdiv{School of Computing}, \orgname{University of Nebraska-Lincoln}, \orgaddress{\city{Lincoln},  \state{NE} \postcode{68588}, \country{USA}}}


\abstract{\textcolor{black}{For next-generation wireless networks, both the upper mid-band and terahertz spectra are gaining global attention. This article provides an in-depth analysis of recent regulatory rulings and spectrum preferences issued by international standard bodies. We highlight promising frequency bands earmarked for 6G and beyond, and offer examples that illuminate the passive service protections and spectrum feasibility for coexistence between terrestrial and non-terrestrial wireless networks.}}



\maketitle

\section{Introduction}
\label{sec:intro}
As wireless communication technologies evolve toward 6G and beyond, previously
underexplored frequency bands have emerged as promising candidates to enable
ultra-high-speed data transmission, ultra-low latency, and next-generation
networking paradigms. Given the increasing demand for higher bandwidth and
emerging applications such as direct-to-device (D2D) satellite service, other
non-terrestrial networks (NTNs), terahertz backhaul, chip-to-chip
communications, and wireless fiber replacement, it becomes apparent that
understanding spectrum availability and regulatory constraints is critical for
product planning by the industry, as well as network build-out strategies for
service providers. This article provides a comprehensive survey of global
spectrum allocation activities through analyses of current regulatory
frameworks, international policies, as well as suitability for network operators
in the future.  Here, we examine allocations and restrictions set by major
regulatory bodies, including the International Telecommunication Union (ITU),
the Federal Communications Commission (FCC), and national telecommunications
agencies across Europe, Asia, North America, and emerging markets.

\subsection{New Spectrum Released in Each Generation}
Since the founding of the cellular telephone industry, new spectrum has been released to meet the growing demands of wireless services and applications. The global community identifies frequency bands that are studied and discussed at the International Telecommunication Union (ITU) World Radiocommunication Conference (WRC) every four years. In the 1970s and 1980s, the first-generation wireless systems around the world were deployed in the 800 and 900 MHz bands. In the second generation (2G) of cellphone technology, the Global System for Mobile Communications (GSM) evolved out of the first pan-European spectrum and technology standardization effort, where countries throughout Europe agreed on a common cellphone spectrum allocation across country borders and on a single GSM air interface. Through ITU, governments around the world agreed upon spectrum allocations for the budding cellphone industry at both 800--900 MHz as well as at 1900 MHz.  In the U.S. and Japan, numerous types of digital standards proliferated, including IS-136 (TDMA), IS-95 (CDMA), and Japan Digital Cellular (JDC)~\cite{rappaport2024wireless}. 

In the 2000s, when the third-generation cellular networks began to roll out with high-speed digital cellular networks, another new frequency band at 2.1 GHz in the U.S. was approved for major standards, such as UMTS and CDMA2000, which provided a much-improved link throughput as packet data was introduced. Other countries began authorizing new spectrum, best suited to dovetail with their own incumbent users, to foster the remarkable growth of cellular telephones. In the 2010s, the fourth-generation (4G) long-term evolution (LTE) technology was adopted as the single, unifying technology standard embraced by global carriers and manufacturers,  leading to incredible market scale and cost savings to consumers. The ubiquity of wireless and the advanced capabilities of LTE required many more spectrum bands to unleash the multimedia and Internet Protocol (IP)-based connectivity. The new frequency bands that support time-division duplex and frequency-division duplex as well as concatenation of bands, including 700 MHz, 1700 MHz, 2300 MHz, 2600 MHz, and 3.5 GHz, respectively~\cite{calabrese2011lte}. 

At the dawn of the 5G era, the FCC led the world in opening up the millimeter wave  (mmWave) spectrum above 24 GHz~\cite{FCC_24GHz_2016, FCC_24GHz_2017}, with spectrum auctions taking place in 2019 for the greatest amount of wireless spectrum ever offered to the industry~\cite{auction2019}. Despite the common misconception that 5G mmWave has not been successful, wireless carriers in India and the USA have reported big rollouts and massive profits using mmWave for fixed wireless access (FWA) and in urban centers and entertainment venues for new customer additions in the 5G mmWave era~\cite{india5G2024,Poggianti2024quiet,Wyrzykowski20245G}.

\begin{table}[h]
    \centering
    \caption{A Summary of New Bands Released in Each Generation of Cellular Networks.}
    \begin{tabularx}{\linewidth}{L|L|L}
    \hline
       \textbf{Generation}  & \textbf{Newly Released Bands Compared to Previous Generation} & \textbf{Total Bandwidth} \\\hline
       1G (1970s and 1980s) & 800--900 MHz (AMPS) &  10 MHz \\\hline
       2G (1990s)  & 800--900 MHz (GSM) and 1900 MHz (GSM \& CDMA) & 160 MHz (GSM) and 170 MHz (CDMA)  \\\hline
       3G (2000s) & 2100 MHz (IMT-2000) & about 320 MHz  \\\hline
       4G (2010s) & 700 MHz, 1700 MHz, 2300 MHz, 2600 MHz, and 3.5 GHz & about 800 MHz  \\\hline
       5G (2020s) & mmWave (24-47 GHz), $>$95 GHz (experimental) & about 1--3 GHz  \\\hline
    \end{tabularx}
    \label{tab:generations}
\end{table}

In March 2019, the FCC opened up spectrum bands above 95 GHz
(Spectrum Horizons, ET Docket 18-21) for unlicensed uses in 95 GHz--3
THz~\cite{rappaport2019wireless}. Today, the Qualcomm Snapdragon X75 modem
supports over 25 different cellphone radio bands from around 400 MHz to 39
GHz~\cite{qualcomm}. The growth in authorized spectrum bands and total
bandwidths enabled by these frequencies is listed in
Table~\ref{tab:generations}.

\subsection{Interest in the Upper Mid-band Spectrum}
\label{sec:upperMidBand}
The upper mid-band spectrum, also known as FR3 (7.125 GHz--24.25 GHz) in 3GPP,
has garnered significant global interest~\cite{bazzi2025upper}. Agencies and
institutes around the world, such as the ITU, the U.S.  National
Telecommunications and Information Administration (NTIA), FCC, European
Telecommunications Standards Institute (ETSI), and Ofcom in the U.K., have been
considering several frequency bands within this spectrum for 5G-Advanced (3GPP
Release 18) and 6G (3GPP Release 19) applications~\cite
{shakya2024comprehensive,ghosh2025new}. The primary interest lies in the
so-called \textcolor{black}{``golden band'' in 7.125--24.25 GHz due to its balance of coverage and
system capacity while supporting wider bandwidths than previous sub-6 GHz
spectrum allocations.} Historically, the lower frequencies, e.g., below 6 GHz,
have been extremely desirable due to their better propagation through buildings
and foliage, although the higher millimeter wave bands offer much wider
allocations of spectrum as well as much greater channel gain due to compact yet
very directional beamforming antennas that overcome the omnidirectional free
space path loss that naturally occurs at higher frequencies
~\cite{5gamericas20246g, rappaport2012cellular, nie201372}.

Another service trend in wireless networking in the upper mid-band spectrum is
the integrated terrestrial and non-terrestrial networks, where satellites can
connect directly to mobile devices, thereby bypassing ground infrastructure
altogether~\cite{azari2022evolution}. Such D2D satellite-to-mobile links can
provide coverage for areas where ground infrastructure is inaccessible or too
expensive to build, such as in rural areas, mountainous areas, or the sea. This
is particularly important given that oceans make up 75\% of land mass on earth,
and rural or mountainous locations are difficult and expensive to serve with
conventional terrestrial infrastructure.  Here, we provide an in-depth
discussion on the new FR3 bands and other promising future spectrum allocations
at the ultra-high frequency (UHF, between 300 MHz and 3 GHz) and mid-band (up to
6 GHz) in Sec.~\ref{sec:below100}.




\subsection{Issues with Current Global Spectrum Allocations in Fixed, Mobile, Defense, and Satellite Services}
Current spectrum allocations adopted by countries often vary widely, leading to compatibility issues and difficulties in harmonizing spectrum allocations across the world.  For example, the 3.5 GHz band in the U.S. is for Citizens Broadband Radio Service (CBRS) where organizations can adopt and build private 5G networks with a shared licensing model~\cite{sohul2015spectrum}. However, in Europe and some countries in Asia, this band is for standard 5G cellular networks. Thus, when equipment providers, such as Ericsson or Nokia, wish to deploy their network equipment in European and U.S. markets, different hardware and firmware must be configured to meet specific regulations in respective regions. For a 5G user who travels between the U.S. and Europe, these variations in spectrum allocations cause product rollout delays and may lead to reduced coverage in certain areas due to the differences in assigned cellular bands and lagging support in devices. Some users may find their cellphones with channels that are operable in one region but not in another, which may lead to degradation in user experience. 


\subsection{\textcolor{black}{Organization and Notations}}
Some recent works have explored the spectrum opportunities~\cite{katwe2024cmwave, rappaport2019wireless,akyildiz2022terahertz, polese2023coexistence}, and this article is unique in that it provides many important recent developments in the rapidly changing field of spectrum allocations, as well as particular insights into the rapidly growing adoption of satellites as part of regional and global cellphone networks. \textcolor{black}{The article is organized as shown in the flowchart in Fig.~\ref{fig:flowchart}}. For the rest of this article, we first present cellular spectrum opportunities
at the desired lower frequencies at UHF, mid-band, and upper mid-band in
Sec.~\ref{sec:below100}. An overview of current spectral allocations across
major regulatory organizations at UHF, mid-band, and upper mid-band for D2D
satellite service is presented in Sec.~\ref{sec:D2D}, followed by a review of
recent experiments in the upper mid-band for conventional cellular services in
Sec.~\ref{sec:experiments_uppermid}. For 5G-Advanced/6G networks, the global
spectrum regulations above 100 GHz are provided in~\ref{sec:regulations}. Then,
a discussion on promising frequency bands above 100 GHz is presented in
Sec.~\ref{sec:potentialBands}, followed by an outlook of research directions for
emerging spectrum allocations in Sec.~\ref{sec:openProblems}. \textcolor{black}{The notations and abbreviations can be found in Table~\ref{tab:notations} on the following page.}

\begin{table}[!h]
    \centering
     \caption{\textcolor{black}{A Summary of Terminology Used in This Paper.}}
    \begin{tabular}{c|c}
    \hline
        Notation and Abbreviation & Definition \\\hline
        AI & Artificial Intelligence \\\hline
        AMPS & Advanced Mobile Phone System \\\hline
        CBRS & Citizens Broadband Radio Service \\\hline
        CDMA & Code-Division Multiple Access \\\hline
        DSA & Dynamic Spectrum Access \\\hline
         D2D & Direct-to-Device \\\hline
         EESS & Earth Exploration-Satellite Service \\\hline
         ETSI & European Telecommunications Standards Institute \\\hline
         EPFD & Equivalent Power-Flux Density \\\hline
         FCC & Federal Communications Commission \\\hline
         FMCW & Frequency-Modulated Continuous Wave \\\hline
         FWA & Fixed Wireless Access \\\hline
         GSM & Global System for Mobile Communications \\\hline
         HAP & High-Altitude Platform \\\hline
         ICT & Information and Communication Technology \\\hline
         IMT & International Mobile Telecommunications \\\hline
         ISM & Industrial, Scientific, and Medical \\\hline
        ITU & International Telecommunication Union \\\hline
        ITU-R WP & ITU Radiocommunication Sector Working Party \\\hline
        IP & Internet Protocol \\\hline
        JDC & Japan Digital Cellular \\\hline
        LEO & Low Earth Orbit \\\hline
        LTE & Long-Term Evolution \\\hline
        LLM & Large Language Model \\\hline
        LoS & Line of Sight \\\hline
        MIMO & Multiple-Input Multiple-Output \\\hline
        ML & Machine Learning \\\hline
        mmWave & Millimeter Wave \\\hline
        MSS & Mobile Satellite Services \\\hline
        NASA & National Aeronautics and Space Administration \\\hline 
        NTIA & National Telecommunications and Information Administration \\\hline
         NTN & Non-Terrestrial Networks \\\hline
         OFDM & Orthogonal Frequency-Division Multiplexing \\\hline
         OOB & Out of Band \\\hline
         RAS & Radio Astronomy Service \\\hline
         RMS & Root Mean Square \\\hline
         TDMA & Time-Division Multiple Access\\\hline
         THz & Terahertz \\\hline
         UAV & Unmanned Aerial Vehicle \\\hline
         UHF & Ultra-High Frequency \\\hline
 WRC & World Radiocommunication Conference \\\hline
 UMTS & Universal Mobile Telecommunications System \\\hline
 3GPP & 3rd Generation Partnership Project \\\hline
    \end{tabular}
    \label{tab:notations}
\end{table}

\begin{figure}[h]
\centering
\begin{tikzpicture}[
  node distance=9mm and 15mm,
  >=Latex,
  mybox/.style={draw, rounded corners, thick, align=left, font=\small,
                fill=blue!5, inner sep=4pt, text width=5.2cm},
  group/.style={draw, rounded corners, thick, inner sep=3pt},
  sub/.style={draw, rounded corners, align=left, font=\footnotesize, fill=white, inner sep=3pt},
  every fit/.style={inner sep=4pt},
  arrow/.style={-Latex, thick}
]

\node[mybox] (S1) {Sec.~1: Introduction};
\node[mybox, below=of S1] (S2) {Sec.~2: Spectrum Opportunities at UHF, Mid-band, and Upper Mid-band for Mobile and Fixed Wireless};
\node[mybox, below=of S2] (S3) {Sec.~3: Spectrum Allocation for D2D Satellite Applications};
\node[mybox, below=of S3] (S4) {Sec.~4: Future Promise of Upper Mid-band for Conventional Cellular Services};
\node[mybox, below=of S4] (S5) {Sec.~5: Global Spectrum Regulations for 5G-Advanced/6G Above 100 GHz};
\node[mybox, below=of S5] (S6) {Sec.~6: Future Frequency Bands and Coexistence Consideration Above 100 GHz};
\node[mybox, below=of S6] (S7) {Sec.~7: Future Research Directions for Emerging Spectrum Allocations and Coexistence};
\node[mybox, below=of S7] (S8) {Sec.~8: Conclusion and Discussion};
\node[sub, right=10mm of S1, yshift=0mm] (S1a) {1.1 New Spectrum Released in Each Generation \\ 
1.2 Interest in the Upper Mid-band Spectrum \\
1.3 Issues with Current Global Spectrum Allocations in\\
Fixed, Mobile, Defense, and Satellite Services};
\node[sub, right=10mm of S3, yshift=0mm] (S3a) {3.1 D2D’s Need for Bandwidth \\ 
3.2 Existing Mobile Satellite Services (MSS) Spectrum \\
3.3 Terrestrial Spectrum under Supplemental Coverage \\from
Space\\
3.4 Prospects of Upper Mid-Band Spectrum for D2D};
\node[sub, right=10mm of S5, yshift=0mm] (S5a) {5.1 International Telecommunications Union (ITU) \\ 
5.2 Federal Communications Commission (FCC) \\
5.3 European Telecommunications Standards Institute \\(ETSI)};
\node[sub, right=10mm of S6, yshift=0mm] (S6a) {6.1 Recent Global Experimental Efforts Above 100 GHz\\ 
6.2 Short-distance “Whisper Radio” at 180 GHz \\
6.3 220 GHz\\
6.4 280 GHz};
\node[sub, right=10mm of S7, yshift=0mm] (S7a) {7.1 New Scenario in the Upper Mid-band\\ 
7.2 AI-assisted Spectrum Allocation Strategies \\
7.3 Energy-efficient Spectrum Sharing\\
7.4 Integrated Terrestrial and Non-terrestrial Networks\\
7.5 Coexistence of D2D with Ground-based \\Astronomy};

\draw[arrow] (S1) -- (S2);
\draw[arrow] (S2) -- (S3);
\draw[arrow] (S3) -- (S4);
\draw[arrow] (S4) -- (S5);
\draw[arrow] (S5) -- (S6);
\draw[arrow] (S6) -- (S7);
\draw[arrow] (S7) -- (S8);
 \draw[arrow] (S1) -- (S1a.west);
 \draw[arrow] (S3) -- (S3a.west);
 \draw[arrow] (S5) -- (S5a.west);
 \draw[arrow] (S6) -- (S6a.west); 
 \draw[arrow] (S7) -- (S7a.west);
\end{tikzpicture}
\caption{\textcolor{black}{Organization of this paper.}}
\label{fig:flowchart}
\end{figure}

\section{Spectrum Opportunities at UHF, Mid-band, and Upper Mid-band for Mobile and Fixed Wireless}
\label{sec:below100}

Frequencies at sub-6 GHz and upper mid-band (7--24 GHz) provide opportunities for burgeoning markets in 5G-Advanced (3GPP Release 18) and future 6G (3GPP Release 19) applications due to their balance of coverage and system capacity. AT\&T recently proposed to the FCC to consider moving the CBRS band out of the current allocation of 3.5--3.7 GHz, and down to 3.1--3.3 GHz in order to relocate 150 MHz of prime cellular spectrum at 3.55 GHz for the cellphone industry~\cite{ATT2024CBRS}. This proposal stems from an observation of a low usage of the CBRS band by the cable industry in the U.S. Similarly, the U.S. Department of Defense (DoD) recently proposed to re-allocate the original CBRS band at 3.5 GHz for 5G services~\cite{DoD2025Spectrum}. The DoD proposal went further, also calling for the release of 50 MHz of new spectrum at 1.3 GHz, 70 MHz at 1.8 GHz, 75 MHz in the upper 5-GHz band, and 125 MHz in the 7-GHz band, respectively~\cite{DoD2025Spectrum}. Among all upper mid-band candidates, 7.125--13.25 GHz is considered the most promising ``golden band'' due to favorable propagation and wider spectrum chunks without adjacent incumbents~\cite{5gamericas20246g, shakya2024comprehensive}. 

The NTIA in the U.S. published an implementation plan in 2024 for the frequency
band in 7.125--8.4 GHz as part of the National Spectrum Strategy, which is
currently under study by federal agencies as
``Strategic Objective 1.2: Ensure spectrum resources are available to support
private sector innovation now and into the
future''~\cite{ntia2024plan}. Meanwhile, the FCC also expanded 12.2--12.7 GHz
for flexible use in terrestrial fixed links and 12.7--13.25
GHz for mobile broadband in 5G in 2023~\cite{12GHz2023}. Similarly, Europe is
considering 7.125--8.4 GHz and 14.8--15.35 GHz for international mobile
telecommunications (IMT) in terrestrial networks toward
6G~\cite{qualcomm2024}. Ofcom in the U.K. has been considering hybrid approaches
to share the 6-GHz band between Wi-Fi and cellphone users~\cite{ofcom2024}. A
more detailed breakdown of the global allocation of the upper mid-band spectrum
is given in Table~\ref{tab:below100}. Another two promising bands are 18.1--18.6
GHz (for expanded satellite operations) and 37.0–37.6 GHz
(for shared usage among federal and non-federal satellite and terrestrial operations),
which are part of the National Spectrum Strategy currently being studied by
NTIA, NASA, DoD, and FCC~\cite{ntia2024plan}.

\begin{table}[h]
    \centering
    \caption{Overview of Allocation and Usage of Upper Mid-band Spectrum in 7--24 GHz.}
    \begin{tabularx}{\linewidth}{L|L|L}
    \hline
       \textbf{Frequency Range}  & \textbf{Primary Usage} & \textbf{Current Global Allocation}  \\\hline
       7.125–8.4 GHz  &  Under study for IMT (5G/6G)~\cite{ntia2024plan} & WRC-27 Agenda Item 1.7 for mobile broadband. Currently used for FSS and FS.\\\hline
       8.4–9.3 GHz  & EESS and military radar (X-band)	& Widely used for space science and defense\\\hline
       10–10.5 GHz (3-cm band) &	Radiolocation service, IMT (under study) & Some interest in mobile broadband (6G), amateur radio and satellite use on secondary basis.\\\hline
       12.2–12.7 GHz &	Fixed Service (FS), Broadcasting Satellite Service &	FCC considering expanding flexible use\\\hline
       12.7–13.25 GHz & Currently limited federal use by NASA	& FCC considering for mobile broadband or expanded use\\\hline
       14–14.5 GHz	& Fixed Satellite Service (FSS) uplinks & GEO and LEO communications (e.g., Starlink uplink)\\\hline
       17.3–17.8 GHz &	Broadcasting-Satellite Service (BSS) & FCC allows non-geostationary  FSS (Space-to-Earth) operations\\\hline
       18.1–20.2 GHz	& FSS&	High-throughput satellites (HTS) and broadband satellites,18.1--18.6 GHz under study by NTIA and NASA (Strategic Objective 1.2)~\cite{ntia2024plan}  \\\hline
       24 GHz & 5G cellular & 5G NR (n258, n259, n260 bands)\\\hline
    \end{tabularx}
    \label{tab:below100}
\end{table}

\section{Spectrum Allocation for D2D Satellite Applications}
\label{sec:D2D}
D2D initiatives seek to extend connectivity to remote areas lacking traditional
cellular infrastructure~\cite{andrews20246G}. November 2022 saw the launch of
the first D2D activity in the cellphone industry when Apple debuted its
emergency text message services to iPhones in partnership with
Globalstar~\cite{globalstar2024}. Currently, the U.S. service provider T-Mobile
has partnered with SpaceX's Starlink satellite system to provide D2D text
message delivery as well as Wireless Emergency Alerts nationwide with the aim of
eliminating ``mobile dead zones''~\cite{tmobile2025}. Meanwhile, AST SpaceMobile is developing a satellite-based cellular broadband network to provide 5G coverage globally, partnering with U.S. carriers AT\&T and Verizon, among others~\cite{att2024}.

\begin{figure}[t]
  \centering
  \includegraphics[width=0.95\linewidth]{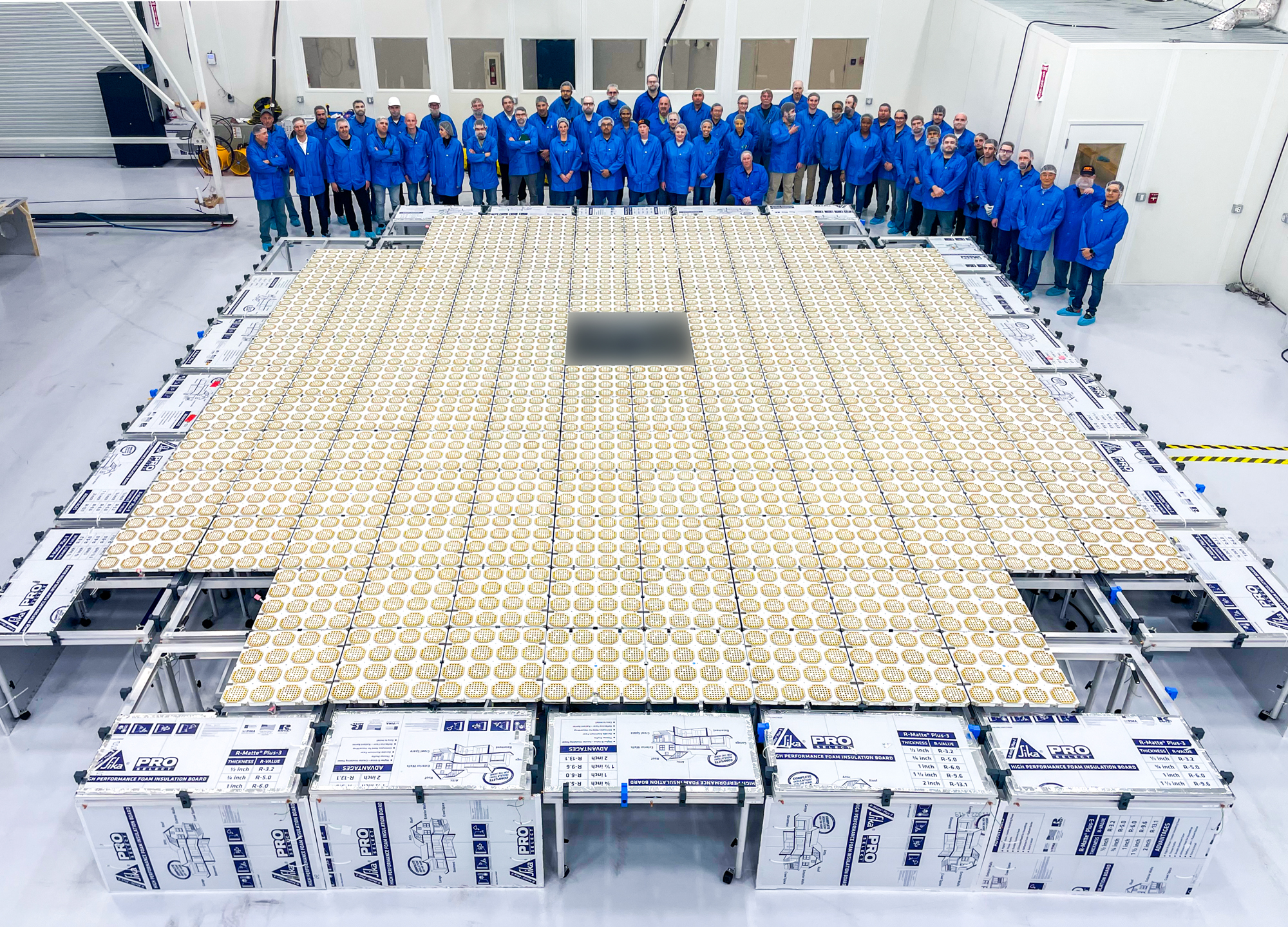}
  \caption{AST SpaceMobile’s 64-m$^2$ BlueWalker 3 test satellite before launch
    in 2022.  The Block 2 BlueBird, scheduled for launches beginning in July
    2025, is three times larger. Image source: AST SpaceMobile.}
  \label{fig:asts-antenna}
\end{figure}

\subsection{D2D's Need for Bandwidth}
D2D initiatives face a key challenge in securing enough bandwidth to support
ambitious proposals for broadband service from space.  Consider AST SpaceMobile
(AST), whose Block 2 BlueBird satellites are each designed to support 2500
adjustable antenna beams, with every downlink beam supporting a bandwidth over
40 MHz in the UHF and L bands, and possibly later the S band
\cite{astArchivedPressRelease2024, kookReportOnAst2024}.  Straightforward
downlink calculations illustrate why ample bandwidth is essential to AST's
broadband ambitions~\cite{andrews20246G}.  Assuming a wavelength $\lambda = 34$ cm (a satellite downlink frequency of 880 MHz in the UHF cellular band), a space-based antenna
area $A = 223$ m$^2$ and efficiency $\eta = 0.7$, the Block 2 phased array gain
is approximately $G_\text{t} = 4 \pi \eta A/\lambda^2 = 42$ dBi, for a
half-power beamwidth of $\theta_\text{t} = 0.028$ rad, or 1.6 deg.  At an
altitude of 725 km ~\cite{boumard2023technical}, the satellite produces a beam
whose narrowest footprint has a 20.3 km diameter and 324 km$^2$ area.  Assuming
a spectral efficiency of 3 bps/Hz (consistent with early testing), a 40-MHz beam
could support a total downlink rate of 120 Mbps.  If directed at a sparse rural
area with population density of 30 users per km$^2$ and a conservative 50\%
smartphone ownership (the U.S. average in 2023 was 90\%), a single beam would
encompass $324 \times 30 \times 0.5 = 4860$ smartphones in its footprint.
Assuming 5\% peak concurrency usage, about 240 of these phones would be active
during peak demand hours, for an equal-division allocation of 500 kbps per
user---far from broadband rates.


To approach scalable broadband, AST and other satellite carriers must either increase their system's area
spectral efficiency (bits/Hz/area) or expand their bandwidth.  The former will be
difficult due to launch vehicle and power limitations.  Narrowing the antenna beamwidth, which occurs as the antenna area increases, appears impractical.  Block 2 BlueBird satellites are already so large
(three times larger than the array shown in Fig. \ref{fig:asts-antenna}) that
only eight fit within the enormous payload fairing of Blue Origin's New Glenn
launch vehicle, the largest for which the Block 2 is
manifest~\cite{harrison2024building}.  A further size increase in pursuit of
higher area spectral efficiency would markedly raise launch costs, eroding the
constellation's economic viability.  Similarly, the Block 2 satellites are
already designed to radiate power at the limits of SCS out-of-band (OOB)
emission and interference restrictions, though a recent loosening of these
limits in the U.S. for Starlink may also be extended to AST, providing
some additional headroom \cite{starlink-waiver-2025}.

Given that AST's Block 2 satellites will already operate near the practical
limits of antenna size and radiated power, expanded bandwidth appears to be the
most viable path to increased D2D throughput. Another approach
  would be to increase the operating carrier frequency since path loss is
  dramatically decreased with increased frequency and fixed antenna area (See
  Sec.~\ref{sec:potentialBands}); however, the tropospheric attenuation and rain
  attenuation would tend to offset such link budget gains at higher
  frequencies~\cite{rappaport2019wireless, xing2021terahertz}. A similar
analysis---not presented here, but yielding similar results---can be carried out
for SpaceX's D2D offering via Starlink to appreciate their need for greater
bandwidth.  The following subsections assess the various possible sources of D2D
spectrum.

\subsection{Existing Mobile Satellite Services (MSS) Spectrum}
Historically, mobile satellite services (MSS) have been assigned to spectrum
bands in L-band (1--2 GHz) and S-band (2--4 GHz)~\cite{doc2021manual}, as well
as slices of the Ku- and Ka-band~\cite{chini2010survey}. One of the advantages
of the existing MSS spectrum is that it is well-coordinated globally and
protected from terrestrial interference. Resolutions from WRC-23 also updated
the Global Maritime Distress and Safety System to strengthen safety and
reliability of MSS communications. These benefits make conventional MSS spectrum
well-suited for D2D, including mission-critical applications, such as
ship-to-shore alerting and connectivity recovery from natural disasters.

The primary impediment to expanded use of MSS for D2D is lack of unencumbered
bandwidth~\cite{musey2025spectrum}. Even existing allocations are typically too
narrow for broadband D2D; e.g., Inmarsat's L-band Broadband Global Area Network
constrains throughput to an average of 650~kbps per channel~\cite{inmarsat2024}.
Against this backdrop, AST's recent purchase of usage rights to 40 MHz of
Ligado's MSS L-band spectrum \cite{ligado-asts-2025} is a rare and impressive
attainment.


\subsection{Terrestrial Spectrum under Supplemental Coverage from Space}
In February 2024, the FCC introduced a groundbreaking regulatory framework
called Supplemental Coverage from Space (SCS) in which spectrum owned by
terrestrial network operators can be used for D2D.  The ultimate goal of SCS is
a ``Single Network Future'' with seamless collaboration between satellite system
operators and terrestrial mobile service providers~\cite{SCS2024}. Under SCS,
two primary approaches are being explored to enable D2D using terrestrial mobile
spectrum: (1) dedicate spectrum solely for D2D service, and (2) share the same
spectrum between terrestrial- and satellite-based
services~\cite{musey2025spectrum}. Here we discuss the benefits and challenges
of both approaches. 

\subsubsection{Dedicated Spectrum for D2D}
In this approach, a terrestrial mobile operator designates specific spectrum
blocks for exclusive use by a satellite partner. \textcolor{black}{For example, T-Mobile and
SpaceX use the 1910--1915 MHz and 1990--1995 MHz bands (i.e., the PCS G Block)
exclusively for D2D services~\cite{PCSG2025}. A dedicated approach simplifies
interference coordination within a carrier's own network and guarantees
consistent availability for satellite services. }Another advantage is the
optimization of waveform and medium access protocols for D2D links, which
improves link reliability and efficiency. However, a dedicated-spectrum model
is inevitably spectrum-inefficient for the carrier---especially in regions where
terrestrial coverage is already adequate---since, due to greater path loss, D2D
has much lower spectral efficiency than the terrestrial network.
  
\subsubsection{Flexible Terrestrial Spectrum Sharing for Both Terrestrial and D2D Services} 
A more flexible alternative allows the mobile spectrum to be exploited both by
the incumbent wireless carriers offering standard cellular service over
terrestrial networks and by satellite networks offering a D2D service to mobile
customers. \textcolor{black}{For Lynk Global, D2D messaging and voice are in sub‑GHz cellular bands via SCS, including a 2025 FCC modification enabling commercial SCS in Guam/Northern Mariana Islands (845.1--845.3 MHz in uplink and 890.1--890.3 MHz in downlink) with partner NTT Docomo Pacific~\cite{FCC_DA_25_385_Lynk_SCS_Guam_CNMI}. } Appropriate technical mechanisms will be required for coexistence
(e.g., time or geographic division, beamforming, and transmitted power control,
among others). The primary benefit of the spectrum sharing model is spectrum
(re-)utilization and reducing barriers for scaling D2D globally. However,
daunting challenges exist in dealing with terrestrial-to-satellite core network
latency, wide Doppler swings, and interference across a D2D satellite's spot
beam while maintaining good spectral efficiency. It may be that the resource
orthogonality and expanded guard intervals required to ensure seamless handover
between terrestrial and D2D coverage will inevitably lead to poor spectral
efficiency in the periphery of terrestrial coverage. \textcolor{black}{Existing studies, including~\cite{lagunas20205g, lim2023interference}, provide quantitative analysis on co-channel interference and out-of-band leakage for both downlink and uplink in satellite networks.}


\subsection{Prospects of Upper Mid-Band Spectrum for D2D}
The upper mid-band spectrum for potential D2D services poses both opportunities
and challenges. While the spectral windows available in the upper-mid band range
(shown in Table~\ref{tab:below100}) have the potential to support broadband
satellite services with significant capacity, the feasibility of space-based
D2D in the upper mid-band is constrained by propagation characteristics.
Satellite transmission links in this range suffer higher path loss for fixed
antenna gains (e.g., shrinking antenna area at higher frequencies) and have
poorer diffraction performance as compared to the lower frequency L-band and
S-band MSS allocations, making it difficult to deliver data to handheld devices,
especially on the uplink under non-line-of-sight (NLoS) scenarios. Additionally,
the increased Doppler shift, which is proportional to the carrier frequency, as
well as more stringent requirements on antenna beam pointing, acquisition, and
tracking, also complicate mobility management and waveform design in the upper
mid-band. Therefore, while technically feasible with advanced transceiver
systems, antenna arrays, adaptive beamsteering, and waveform optimization,
space-based D2D in the upper mid-band may be limited to favorable propagation
conditions or may require hybrid architectures with local terrestrial repeaters
or bi-directional beamforming from space and at the mobile device.

Worth noting is that for fixed satellite service (FSS) systems operating around
Ku- and Ka-band, spectrum sharing relies heavily on \emph{spatial multiplexing}
through beamforming at both the satellite and user terminals, such as is
employed by the Starlink Ku-band terrestrial user terminals. But handheld
devices are too small to support tight device-side beamforming.  Thus, their
quasi-omnidirectional antennas will collect signals from large sectors of the
sky. As a result, spectrum sharing through spatial multiplexing from space is
unlikely in D2D applications.

\section{Future Promise of Upper Mid-band for Conventional Cellular Services}
\label{sec:experiments_uppermid}
As the interest in the upper mid-band spectrum grows globally (previously discussed in Sec.~\ref{sec:upperMidBand}), this spectral range, which covers more than 8~GHz of bandwidth~\cite{5gamericas20246g}, promises great potential in achieving a balanced tradeoff between throughput and coverage~\cite{bazzi2025upper}. Only lately has the global wireless community started to develop in-depth knowledge of propagation conditions for mobile applications of this emerging spectrum, as it has traditionally been allocated for satellite, fixed point-to-point, and military applications, where the global cellular network community in 3GPP standard bodies had assumed no difference in radio propagation characteristics between the conventional cellular frequencies below 6 GHz and frequencies up through 24 GHz. 

In recent channel studies at the upper mid-band, various institutions across the world have found that the channel models (such as delay spread and angular spread models) currently assumed by the 3GPP standard bodies are inadequate and require revision. In particular, researchers at NYU WIRELESS studied an indoor factory channel at 6.75
GHz and 16.95 GHz with a bandwidth of 1 GHz~\cite{ying2025upper}. The work demonstrates a frequency-dependent behavior of root-mean-square (RMS) delay spread and RMS angular spread. Another study on penetration loss at 6.75 GHz and 16.95 GHz over ten
different building materials and antenna polarizations suggests existing 3GPP material penetration models may be required for revision~\cite{shakya2024wideband}. In addition, another work~\cite{shakya2025urban} conducted in an urban outdoor setting shows that measured mean values of NLOS RMS delay spread and RMS angular spread are consistently lower in New York City compared to 3GPP model predictions using TR 38.901, which brings the knowledge for the first time to the research community. 


Researchers at Seoul National University also studied the upper mid-band in an urban microcellular scenario in Seoul~\cite{yang20256G}. Their work corroborates the findings of NYU in that the 3GPP model for RMS delay spreads in the upper mid-band currently used by 3GPP is much larger than was observed in the streets of Seoul, South Korea. This finding is an important design consideration for the 6G standardization process, since a smaller RMS delay spread implies the physical existence of a larger coherence bandwidth in the channel that will support wider bandwidth, hence leading to more subcarriers in Orthogonal Frequency-Division Multiplexing (OFDM) than at sub-6-GHz frequencies.

A set of outdoor double-directional measurements targeting the urban
microcellular scenario is conducted on the campus of the University of Southern
California using a vector network analyzer-based channel sounder that can sweep
across 6--14 GHz~\cite{abbasi2024ultra}. Key findings from the measurement
campaign also confirm the stronger received power as compared to theoretical
predictions from the Friis path loss model due to multipath components, whereas
angular spreads at the transmitter and receiver 
demonstrate general consistency across different frequencies.

In a multi-institutional collaboration in the U.S., including NYU, the University of Notre Dame,
and Florida International University (FIU), researchers developed a custom software-defined radio (SDR) platform and conducted experiments in urban outdoor settings in a wide spectral range of 6--24 GHz~\cite{kang2024cellular}. The studies, enabled by this SDR platform, allow for flexible waveform generation and reception, as well as adaptive wideband operations in the upper mid-band, thus facilitating channel sounding and interference management in various application scenarios, including the coexistence of satellite and terrestrial cellular networks.

European efforts in the upper mid-band are also gaining traction. For example, Politecnico di Milano investigated
the 6.425--7.125 GHz band using a customized 5G new radio (NR) testbed for urban and factory
automation scenarios~\cite{morini2023will}. Their work provides an evaluation of the upper
mid-band's capabilities in supporting deterministic latency and achieving reliability goals in
Industry 4.0 environments while maintaining adequate coverage. Such an overview of recent efforts by research and industrial groups investigating
the propagation characteristics of the upper mid-band (FR3) is presented in
Table~\ref{tab:upper-midband-groups} on the following page. 

\begin{table}[h]
\centering
\caption{Research and industrial groups conducting upper mid-band (FR3) channel measurements across various global scenarios.}
\begin{tabularx}{\linewidth}{L|L|L|L|L|L}
\hline
\textbf{Institution}  & \textbf{Frequency Range (GHz)} & \textbf{Scenario} & \textbf{Bandwidth (MHz)} & \textbf{Channel Sounder Type} & \textbf{References} \\\hline
NYU WIRELESS & 6.95 \& 16.95 & Indoor factory, indoor office, urban microcell & 1000 & Sliding correlator-based & \cite{ying2025upper,shakya2025urban, shakya2024wideband} \\\hline
Seoul National University & 7.5 GHz & Dense urban & -- & -- &\cite{yang20256G}\\\hline
USC &  6--14 & Urban microcell &  8000 & VNA-based & \cite{abbasi2024ultra} \\\hline
NYU, Notre Dame, FIU, etc.  & 6--24 & Urban microcell & -- & Custom SDR platform & \cite{kang2024cellular} \\\hline
Politecnico di Milano & 6.425--7.125 & Urban microcell, factory automation & 80 & customized 5G NR testbed & \cite{morini2023will} \\\hline
\end{tabularx}
\label{tab:upper-midband-groups}
\end{table}

\section{Global Spectrum Regulations for 5G-Advanced/6G Above 100 GHz}
\label{sec:regulations}
As different countries adopt different strategies in developing 5G and future 6G networks, it is necessary to review current global spectrum regulations above 100 GHz to identify opportunities for future service expansion and coexistence in the 5G-Advanced/6G era. In this section, we present the key agenda items toward  World Radiocommunication Conference 2027 (WRC-27) and offer perspectives indicated by major standard bodies, such as ITU, FCC, and ETSI. 

\subsection{International Telecommunications Union (ITU)}
The ITU is an agency within the United Nations that oversees global telecommunications, information, and communication technology (ICT) policies. The ITU functions to regulate and harmonize global spectrum usage, set international standards, facilitate international coordination and cooperation, as well as promote global connectivity. ITU also organizes world radiocommunication conferences (WRCs) every four years to lay out agenda items for discussion by participants from global industry, governments, and research groups, which shape the prospects of the wireless industry. One of the agenda items in WRC-27 focuses on evaluating potential regulatory measures to protect passive services operating in frequency bands above 76 GHz (Agenda Item 1.18)~\cite{itu1.18}. Specifically, Agenda Item 1.18 seeks to ensure the global protection of the Earth Exploration-Satellite Service (EESS) (passive) and the Radio Astronomy Service (RAS) from unwanted emissions originating from active services. A study is being conducted under the guidance of Resolution 712 (WRC-23), which is an in-depth assessment of the interference risks posed by emerging active radio systems~\cite{wrc2023itu}. Given the increasing interest in utilizing the mmWave and sub-THz spectrum for advanced wireless communications, including 6G and beyond, the need for spectrum coexistence analysis and interference mitigation strategies is becoming more pressing.

The ITU Radiocommunication Sector (ITU-R) Working Party (WP) 7C leads discussions on remote sensing services, which include both active and passive Earth observation systems, such as the EESS. 
In preparation for WRC-27 Agenda Item 1.18, WP 7C has developed a draft framework for conducting spectrum sharing and compatibility studies related to EESS operations in frequency bands above 76 GHz. The framework lays the foundation for evaluating potential regulatory and technical measures necessary to protect passive sensing applications~\cite{de2020agenda}. One of the key aspects of the current ITU-R studies conducted by the WP 7C is the assessment of unwanted emissions from active services, particularly those from high-frequency communication and sensing technologies. The reason is that with the highest frequency ranges considered in the current draft framework for radiolocation service in WRC-27 include 231.5--275 GHz and 275--700 GHz, some have already been allocated or identified for EESS, RAS, fixed and mobile services, among other applications~\cite{united2024code}. Within the working party, involvement of major stakeholders such as the National Aeronautics and Space Administration (NASA) and Amazon underscores the strategic importance of these interference risks and coexistence studies, as both organizations have vested and sometimes contradictory interests in space-based and terrestrial applications that rely on access to mmWave and THz spectra.

\begin{figure*}
    \centering
    \includegraphics[width=0.95\linewidth]{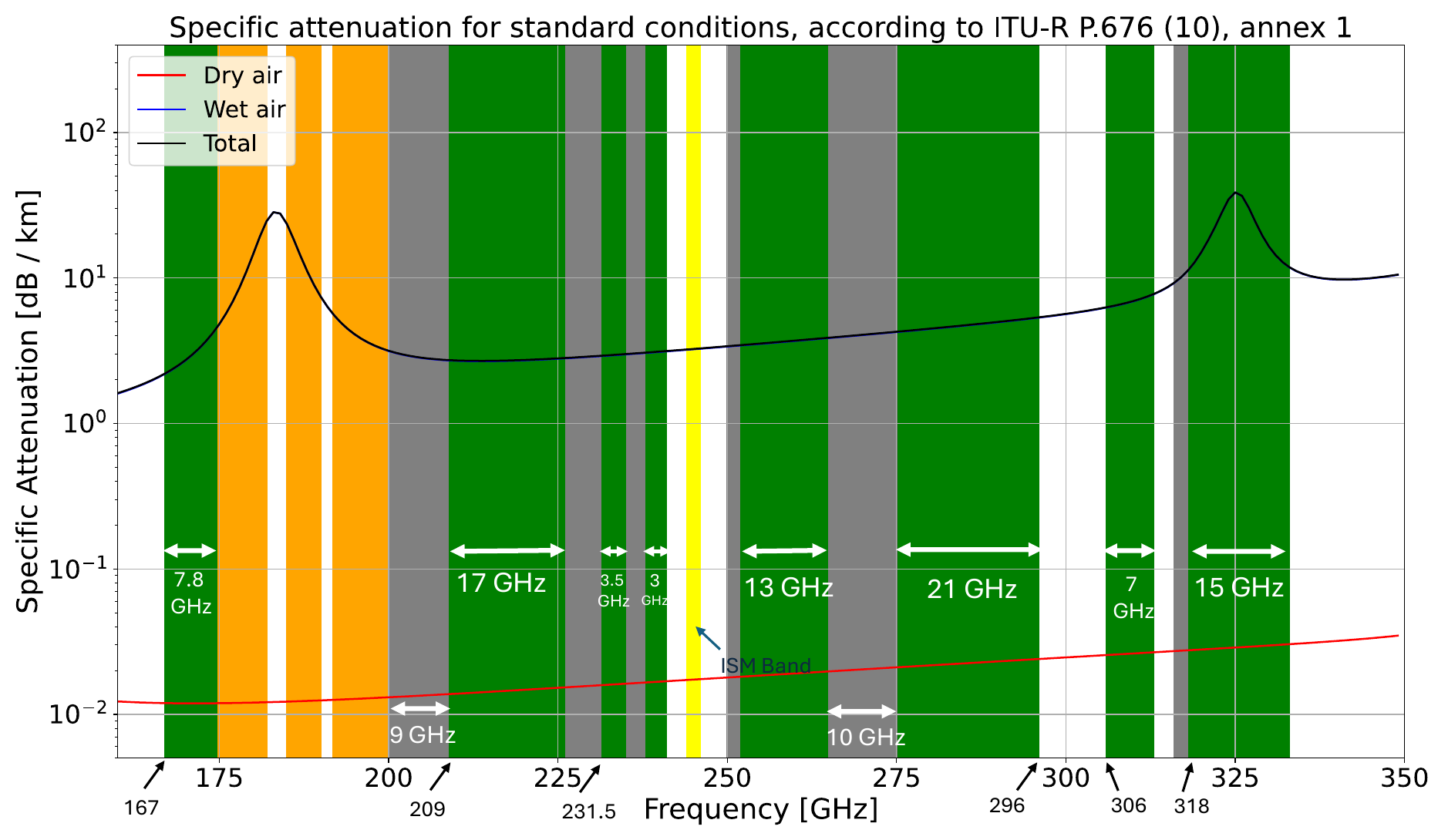}
    \caption{Spectrum allocation from 170 GHz to 350 GHz regulated by ITU. Green-colored bands are for fixed or land mobile services; gray-colored bands are RR5.340 prohibited bands or reserved for passive services, including EESS; orange-colored bands are for inter-satellite services; and the yellow-colored band is for the ISM band at 244 GHz~\cite{united2024code, rappaport2019wireless, polese2023coexistence}. The attenuation traces overlap between wet air (blue curve) and total air (black curve) conditions. }
    \label{fig:allocation}
\end{figure*}

In parallel with the protection of passive services, WRC-27 will also consider the potential identification of several frequency bands for International Mobile Telecommunications (IMT). According to the ITU draft agenda~\cite{itu1.18}, the upcoming WRC-27 will explore the feasibility of allocating specific sub-THz bands for future mobile broadband applications, in accordance with Resolution 255 (WRC-23). The targeted bands under consideration include 102--109.5 GHz, 151.5--164 GHz, 167--174.8 GHz, 209--226 GHz, and 252--275 GHz, all of which exhibit promising characteristics for ultra-high-throughput wireless communications for potential 6G deployments~\cite{wrc2023itu}.  The identification of the candidate frequencies (102--109.5 GHz, 151.5--164 GHz, 167--174.8 GHz, 209--226 GHz, and 252--275 GHz) for IMT use also raises regulatory concerns, mostly on coexistence with passive services. The overlap between proposed IMT bands and existing EESS and RAS allocations (previously demonstrated in~\cite{xing2021terahertz}, also shown in Fig.~\ref{fig:allocation} for portions above 170 GHz) calls for the development of advanced spectrum-sharing frameworks and interference mitigation techniques. A more in-depth analysis of some promising frequency bands within or close to these targeted bands is provided in Sec.~\ref{sec:potentialBands}. 

In a most recent ITU-R Working Party 1A document (Document 1A/88-E) released in May 2025~\cite{japan2025THz}, a revised outline is proposed by Japan on spectrum utilization, channel characteristics, and applications for terrestrial active services, such as fixed, mobile, radiolocation, amateur, and Industrial, Scientific, and Medical (ISM) from 275 GHz to 1 THz. In particular, new use cases are being highlighted, which include 1) for fixed services: THz multi-hop mesh backhaul and multi-antenna spatial-multiplex links, 2) for mobile services: in-cabin THz hotspots, train-to-train links, augmented reality/virtual reality applications, THz device-to-device communications at home,  and THz multi-hop sidelink relays. A new radiolocation use case is also proposed, which is a walk-through security imaging system co-located with THz ``Touch-and-Get'' data kiosks.
A new use case proposed for the ISM bands is a high-/low-power continuous-wave illumination imaging system and frequency modulated continuous wave (FMCW) radars for standoff detection and non-destructive testing. These new use cases are expected to generate more substantial system parameters (e.g., power, antenna gain, modulation scheme, bandwidth) to be further studied under simulation or measurements and researched for compatibility with passive services.

\subsection{Federal Communications Commission (FCC)}

The FCC serves as the primary regulatory body in the U.S. that is responsible for managing spectrum resources across various applications, including wireless communications, passive sensing, and experimental radio services. In the U.S. Code of Federal Regulations (CFR), Title 47 provides the legal framework for spectrum allocation and regulation, particularly in Part 2 (Frequency Allocations and Radio Treaty Matters; General Rules and Regulations) and Part 5 (Experimental Radio Service)~\cite{united2024code}. \textcolor{black}{The FCC Office of Engineering and Technology also provides guidelines in radiated emission measurement for devices operating above 95 GHz up to 750 GHz~\cite{FCC_OET_2025_SubTHzGuidance}.} These regulations ensure a structured approach to spectrum access while accommodating scientific, industrial, and communication needs. In general, the spectrum bands can be categorized into prohibited bands reserved for remote sensing purposes and frequency bands for fixed, mobile, and other applications.

\subsubsection{Prohibited Bands for Passive Sensing and Earth Atmospheric Observations}

To protect the integrity of ground-based atmospheric sensing and passive scientific applications, certain frequency bands have been prohibited from active transmissions. According to FCC regulations, the bands required to protect passive Earth observation and atmospheric monitoring systems include 200--209 GHz, 226--231.5 GHz, and 250--252 GHz~\cite{united2024code}. These prohibited bands are crucial for applications including radio astronomy, climate monitoring, and remote sensing, where even small interference could result in observation inaccuracy. These prohibited bands align with those of ITU to preserve passive sensing capabilities, particularly in mmWave and sub-THz frequency ranges. Similar to the U.S., in Europe, the U.K., Japan, and Australia, these three bands (200--209 GHz, 226--231.5 GHz, and 250--252 GHz) are also reserved for passive sensing and radio astronomy.

\subsubsection{Spectrum for Fixed and Mobile Services}

The National Telecommunications and Information Administration (NTIA) within the U.S. Department of Commerce, in conjunction with the FCC, has identified several high-frequency bands for fixed and mobile service applications, which permit active communication applications under defined conditions. These bands include 209--226 GHz, 231.5--235 GHz, 238--241 GHz, and 252--265 GHz~\cite{united2024code}. These sub-THz allocations provide significant opportunities for high-frequency communication technologies, including 6G mobile and fixed wireless networks, ultra-high-speed backhaul, and point-to-point wireless links. In addition, several frequency bands, including 296--306 GHz, 313--318 GHz, and 333--356 GHz, can also be used for similar mobile and fixed wireless services under the condition of necessary protection for passive sensing and radio astronomy. Additionally, the 244--246 GHz band has been designated for ISM equipment, with a center frequency at 245 GHz~\cite{ISM2019}. This allocation supports applications such as high-frequency industrial processes, biomedical research, and precision sensing systems. 

\subsubsection{Earth Exploration-Satellite Service (EESS)}
The EESS plays a major role in remote sensing applications, including climate monitoring, atmospheric studies, and Earth observation. Specific frequency bands within the 275--1000 GHz range have been identified for EESS operations, as listed in ITU-R Radio Regulations (RR) No. 5.565, which includes bands such as 275--277 GHz, 294--306 GHz, 316--334 GHz, and 342--349 GHz, among others~\cite{united2024code}. These allocations facilitate high-precision sensing of atmospheric, oceanic, and land-based environmental parameters. According to ITU-R Report SM.2450 (2019) and related studies~\cite{itu2450}, it has been determined that no additional regulatory protection measures are required for EESS operations in the 275–296 GHz, 306–313 GHz, and 320–330 GHz bands, provided that active systems comply with operational parameters specified in ITU-R recommendations. This suggestion deems the coexistence between EESS and other services in these bands as feasible, which relieves the concern over potential interference. As THz communication technologies mature, further studies may be necessary to assess the long-term impact on spectral integrity and data quality of EESS.






\subsection{European Telecommunications Standards Institute (ETSI)}

\begin{figure}[ht]
    \centering
    \includegraphics[trim={8mm, 0, 12mm, 15mm},clip, width=1\linewidth]{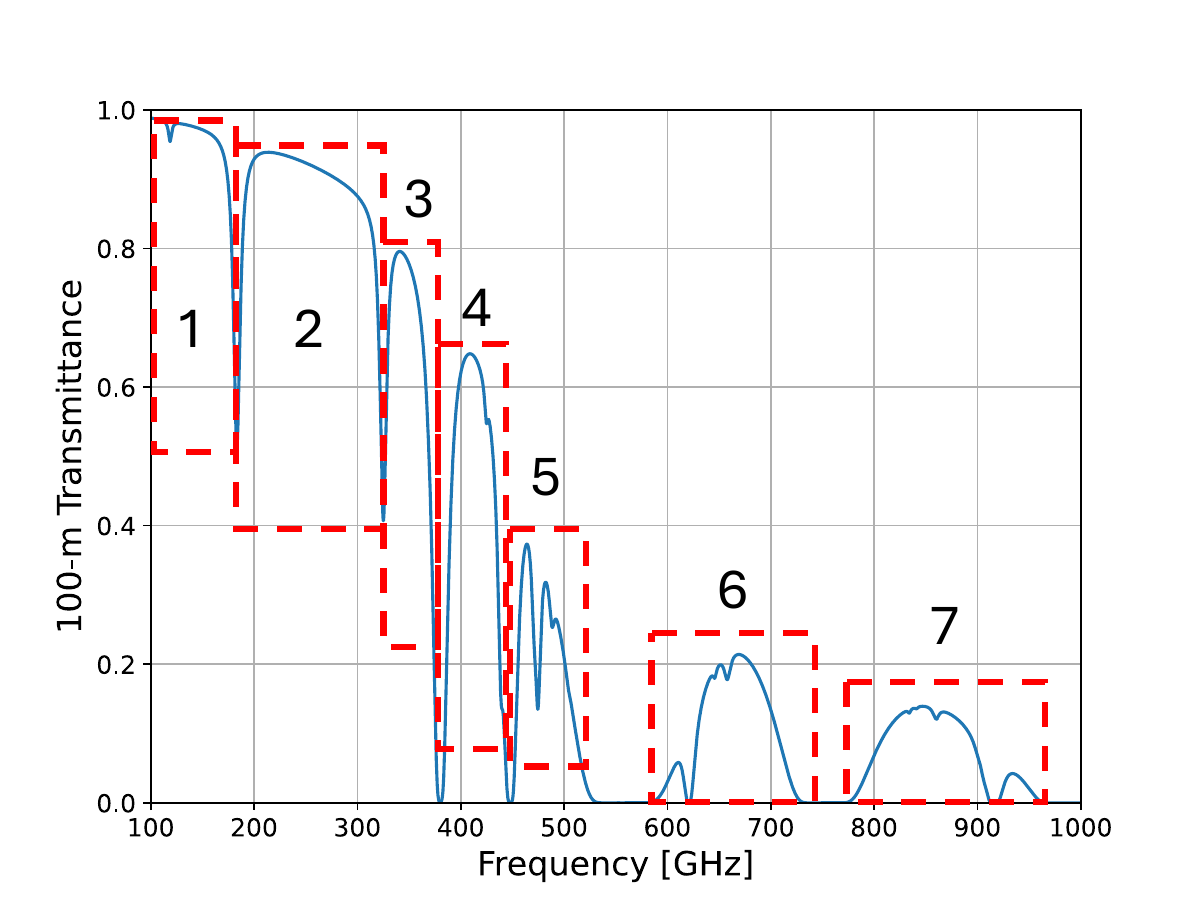}
    \caption{Seven transmission windows defined by ETSI from 100 GHz to 1 THz at the link distance of 100 meters at sea level with a temperature of 15$^\circ$C and water vapor density of 7.5 g/m$^3$~\cite{ETSI2024GR2}.}
    \label{fig:etsi}
\end{figure}

The European Telecommunications Standards Institute (ETSI), especially its Industry Specification Group (ISG) on THz, has conducted extensive studies on the spectrum above 100 GHz, which include promising use cases~\cite{ETSI2024GR1} and feasible frequency bands~\cite{ETSI2024GR2}, as well as extensive survey on propagation channel measurements and developed channel models that can support future high-throughput wireless communication systems~\cite{ETSI2024GR3}. In its research, ETSI categorizes the 100 GHz to 1 THz range into distinct \emph{spectral regions} based on their propagation characteristics, absorption losses, and potential for communication applications, which ETSI names as ``transmission windows'' and identifies specific frequency bands where electromagnetic waves experience lower atmospheric attenuation and can thus enable more efficient data transmission~\cite{ETSI2024GR2}. As shown in Fig.~\ref{fig:etsi}, these transmission windows serve as guidelines for selecting feasible bands for applications such as 6G ultra-high-speed backhaul links, THz high-resolution imaging, and next-generation radar systems, among other communication and sensing applications. ETSI ISG's studies provide an insightful reference in standardizing the use of THz frequencies, which guarantee that future wireless systems can both operate efficiently and coexist in harmony with incumbent services such as passive sensing, radio astronomy, and EESS.








\section{Future Frequency Bands and Coexistence Consideration Above 100 GHz}
\label{sec:potentialBands}

In this section, we discuss frequency bands above 100 GHz that are promising in 6G wireless networks for cellular and inter-satellite services. When evaluating potential frequency bands above 100 GHz, major factors that should be taken into account include the attenuation caused by atmospheric components, including oxygen, water vapor, and other molecular absorptions, according to ITU-R P.676, which provides specific and path gaseous attenuation due to oxygen and water vapor~\cite{ITU-RP676}. The specific attenuation under dry air and wet air (with water vapor content of $7.5~\mathrm{g/m}^3$) in the frequency range of 170--350 GHz is shown in Fig.~\ref{fig:allocation}.

A long-held misconception about frequency bands above 100 GHz with excessively high path loss along propagation channels is that when the free-space path loss grows proportionally with the increase of frequency, the channel becomes extremely lossy. However, this view overlooks the scaling effect of antenna gains, which means that as a steerable or fixed beam maintains the same effective aperture, as the frequency increases, the link has a much reduced loss~\cite{xing2018propagation}. \textcolor{black}{Modern wireless systems in cellular or Wi-Fi networks utilize beamforming to ensure that maximum or near maximum antenna gain is realized, especially at mmWave frequencies.} To further illustrate this effect mathematically, suppose the gain of an antenna is expressed as
\begin{equation}
    G_{\mathrm{TX/RX}} = \frac{4\pi A_e^{\mathrm{TX/RX}}}{\lambda^2},
    \label{eq:gain}
\end{equation} where $A_e^{\mathrm{TX/RX}}$ is the effective aperture of the beam at either the transmitter (TX) or the receiver (RX) and $\lambda$ is the wavelength. Substituting the antenna gain expression into the received power calculation according to the Friis free-space path loss formula (assuming that the transmitted power $P_{\mathrm{TX}}$ and propagation distance $d$ remain fixed), the received power is expressed as
\begin{align}
    P_{\mathrm{RX}} &= P_{\mathrm{TX}}G_{\mathrm{TX}}G_{\mathrm{RX}}\left(\frac{\lambda}{4\pi d}\right)^2 \\\nonumber
    &=P_{\mathrm{TX}}\frac{A_{\mathrm{e}}^{\mathrm{TX}}A_{\mathrm{e}}^{\mathrm{RX}}}{\lambda^2d^2}.
\end{align} This relationship above shows an increased received power (reduced loss)
as frequency increases (wavelength decreases) while fixing the antenna apertures. Extensive field experiments and simulations based on NYUSIM up to 150 GHz in both MATLAB and ns-3 also corroborate these mathematical insights~\cite{ju2023142,poddar2023full,kanhere2025calibration,shakya2024radio}. Additionally, as Eq.~\eqref{eq:gain} shows, the antenna gain scales proportionally with the frequency. Therefore, an antenna array with a fixed aperture provides much higher gain at mmWave and sub-THz frequencies~\cite{rappaport2012cellular,nie201372}. Such an additional gain largely offsets the distance-dependent path loss in free space or lightly/non-obstructed radio channels. Hence, a mmWave and sub-THz link can deliver more received power than a lower‑frequency link~\cite{ju2019scattering}. Mobile carriers are already exploiting this property to roll out FWA services with massive bandwidth in the mmWave bands. 


\subsection{Recent Global Experimental Efforts Above 100 GHz}
A snapshot of the most recent global efforts (results published in the recent two years) in THz channel measurement campaigns is provided in Table~\ref{tab:thz-measurement-groups}, highlighting the diversity of frequency ranges, deployment scenarios, and measurement bandwidths. These efforts span academia, industry, and government laboratories, emphasizing the growing momentum toward establishing realistic channel models and experimental platforms for future THz wireless systems. \textcolor{black}{In the following, we summarize these recent results according to the test frequency and experiment scenario, highlighting key observations.}

\textcolor{black}{At the lower end of the THz spectrum (around 130--150 GHz), NYU WIRELESS has conducted measurements at 142 GHz using a 1 GHz bandwidth sliding correlator-based sounder in a variety of scenarios, such as indoor office, factory, and urban outdoor~\cite{ju2023142,shakya2024radio,kanhere2025calibration}. These studies provide key insight on statistical channel characterization and path loss modeling in industrial and urban outdoor scenarios at various THz bands, highlighting the potential for ultra-broadband communications in the 6G era. A study conducted by Princeton University focuses on diffuse scattering at 140 GHz with 10 GHz of bandwidth, contributing insights into reflection and multipath characteristics in the lower THz regime~\cite{shen2024characterizing}. The National Institute of Standards and Technology (NIST) has developed a testbed operating in the similar band of 141 GHz (in the D-band) with bandwidths up to 6.4 GHz for real-time near-field switch beamforming~\cite{bang2025real}. Notably, this channel sounder achieves a fast channel sweep of 512 $\mu$s and a find 3D spatial resolution of 2~cm. }

\textcolor{black}{Another set of recent work focuses on studying the backhaul feasibility of different THz bands. Researchers from Oklahoma State University demonstrates through extensive field measurements the feasibility of outdoor links over 2.9 km at 130 GHz with a maximum 5.5 GHz bandwidth~\cite{ohara2024long}, highlighting the potential for long-range non-line-of-sight applications. A collaborative effort from Northeastern University, Florida International University, and SUNY Polytechnic Institute developed a software-defined radio (SDR) platform at 147 GHz with the connectivity of a 2-km link~\cite{karunanayake2024pantera}. Institutions in Europe and Asia also contribute significantly to showcase the long-distance connectivity of THz bands. For example, Chalmers University and Ericsson in Sweden conducted W-band measurements (92--114 GHz) over a 1.5-km outdoor link~\cite{horberg2022wband}.}

\textcolor{black}{Moving on to the middle to high end of the THz spectrum (around 200--725 GHz), Collaborative efforts also explore novel scenarios such as chip-to-chip links at 300 GHz (Georgia Institute of  Technology)~\cite{zajic2023future}, jamming and security at 197.5 GHz (Brown and Rice)~\cite{shrestha2022jamming}, and weather-affected propagation, including snowfall, at 140 and 270 GHz (Brown and Beijing Institute of Technology)~\cite{liu2024impact}. Researchers from Shanghai Jiao Tong University in China reported channel behavior in highly directional environments such as L-shaped intersections using bands centered around 300 and 360 GHz~\cite{wang2025300}. Notably, the French National Centre for Scientific Research (CNRS), Lille University, and Osaka University extended measurements into the upper THz spectrum, demonstrating a photonics-enabled link operating from 500--724 GHz with a bandwidth of 250 GHz, achieving throughput of 1.04 Tbps~\cite{ducournau2023photonics}.}

Such a wide range of frequency bands and experimental scenarios being studied thus far reflects both the rising interest and deployment challenges of THz bands. \textcolor{black}{Recent analytical studies  above 100 GHz in satellite networks also show promising results in gradually bridging the gap between theory and reality~\cite{torrens2024modeling, aliaga2022joint,aliaga2023cross, aliaga2023enhancing,aliaga2024analysis, alqaraghuli2023road, alqaraghuli2021performance}.} Standardization efforts (such as IEEE 802.15.3d~\cite{802153d}) are still taking shape to identify feasible spectrum bands, develop accurate THz channel models, novel waveforms, and other physical-layer solutions. Based on the current spectrum allocation regulated by the ITU, several promising frequency bands that may satisfy both throughput and coexistence demands above 100 GHz are discussed below.

\begin{table}[h]
\centering
\caption{Overview of research and industrial groups conducting THz channel measurements.}
\begin{tabularx}{\linewidth}{L|L|L|L|L}
\hline
\textbf{Institution} & \textbf{Frequency Range (GHz)} & \textbf{Scenario} & \textbf{Bandwidth }  & \textbf{References} \\
\hline
NYU WIRELESS & 142 & Indoor office \& factory, urban outdoor & 1 GHz & \cite{ju2023142,shakya2024radio,kanhere2025calibration}  \\\hline
Northeastern University,  FIU, and SUNY-Poly& 147 & Testbed development & 64 MHz & \cite{karunanayake2024pantera}\\\hline
USC and  Samsung Research America & 140 & Urban outdoor & 1 GHz & \cite{molisch2024properties}\\\hline
Oklahoma State University & 130 & Outdoor link at 2.92 km & 5.5 GHz  & \cite{ohara2024long}\\\hline
Georgia Tech & 300 & Chip-to-chip & 12 GHz & \cite{zajic2023future} \\\hline
Brown University and Beijing Institute of Technology & 140 and 270 & Outdoor with snowfall & N/A &\cite{liu2024impact} \\\hline
Brown University and  Rice University & 197.5 & Security, jamming  & 500 MHz & \cite{shrestha2022jamming}\\\hline
Princeton University& 140 & Diffuse scattering & 10 GHz & \cite{shen2024characterizing}\\\hline
NIST & 142 & Testbed development for joint sensing and communication & 6.4 GHz & \cite{bang2025real} \\\hline
Chalmers University of Technology and Ericsson & 92--114 (W-band) & 1.5-km outdoor link & 2& \cite{horberg2022wband} \\\hline
Shanghai Jiao Tong University  &  220, 306--321, and 356--371  & Outdoor and L-shaped intersection & 15 & \cite{wang2025300,li2024220}\\\hline
Technical University  of Braunschweig & 200–300 & Indoor, LoS and NLoS environments & 2–4  & \cite{aldabbagh2025characterization} \\\hline
CNRS, Lille University, and Osaka University  & 500--724 & Measurement system development, testbed with 1.04 Tbps link & 250 GHz & \cite{ducournau2023photonics} \\\hline
\end{tabularx}
\label{tab:thz-measurement-groups}
\end{table}

\subsection{Short-distance ``Whisper Radio'' at 180 GHz}

As shown in Fig.~\ref{fig:allocation}, the wet-air attenuation at around 180 GHz is approximately 20~dB/km, making its usage primary toward short-range communications under wet air conditions. However, this band is favorable when the water attenuation effect is less prominent, which usually occurs above the atmosphere. According to RR 5.562H, the spectral bands in the range of 174.8 GHz--182 GHz and 185 GHz--190 GHz are provisioned for the inter-satellite service for satellites in the geostationary-satellite orbit~\cite[\S4.1.3]{doc2021manual}. According to RR 5.558, a bandwidth of 300 MHz is allocated between 174.5 GHz and 174.8 GHz for fixed, mobile, and inter-satellite services around 180 GHz. In addition, there is 7.5 GHz of bandwidth allocated for this purpose from 167 GHz, amounting to a total of 7.8 GHz feasible for broadband communication purposes. 


\subsection{220 GHz}

The 209--225 GHz band offers a bandwidth of 17 GHz, which is also promising to achieve ultra-high throughput in the THz spectrum. 
This frequency range, which is under study by different research groups~\cite{aldabbagh2025characterization,liao2024measurement, li2024220} in both indoor and outdoor scenarios, has the potential to support high-throughput wireless transmission and short-range THz networks. A portion of this band, specifically 217--226 GHz, is currently allocated for space-based radio astronomy~\cite{united2024code}. However, given the space-borne platform of radio astronomical observations and the directional characteristics of radio telescopes, this 217--226 GHz band may still be potentially reusable for terrestrial fixed and mobile services.

The feasibility of spectrum sharing in this range is further supported by the high atmospheric attenuation at sub-THz frequencies, which inherently limits interference between terrestrial and space-based applications. Additionally, the utilization of dynamic spectrum access (DSA) mechanisms and adaptive beamforming techniques could further enhance coexistence to minimize interference and manage spectrum access. Therefore, the 209--225 GHz band presents an opportunity for advancing future high-speed wireless communications while preserving existing space-based radio astronomy operations. 

%


\subsection{280 GHz}


The 275--296 GHz frequency range provides a contiguous bandwidth of 21 GHz, which sets itself as a promising candidate for future ultra-broadband wireless communication systems. However, there is currently no formal allocation above 275 GHz from either ITU or FCC for communication services, leaving an open regulatory landscape for future standardization and spectrum policy development. Because of the benefit of this spectrum's ultra-wide bandwidth, research groups around the world have been actively conducting experiments in the frequency band above 275 GHz to evaluate its feasibility for practical wireless connectivity. Recent demonstrations, including long-range point-to-point links at 300 GHz~\cite{sen2023multi}, demystify the long-time misconception about the propagation distance limitation in these tremendously high frequencies~\cite{rappaport2019wireless}. More data-hungry applications and services, such as holographic massive multiple-input multiple-output (MIMO) communications, will likely find current networks as a bottleneck and bandwidth allocation limited, necessitating next-generation wireless networks to explore alternative options at THz bands to satisfy these growing demands. 

One of the primary advantages of this band is its favorable propagation characteristics, particularly in low-humidity environments and line-of-sight (LoS) conditions. Compared to other THz bands, the 275--296 GHz range exhibits moderate atmospheric attenuation, which makes it well-suited for applications such as indoor point-to-multipoint communications, ultra-high-speed network-on-chip interconnects, wireless data center, and THz-band fiber replacement technologies. Additionally, the contiguous bandwidth of 21 GHz in this spectrum range enables multi-gigabit per second (Gbps) to terabit per second (Tbps) data transmission, which provides a distinct benefit for high-capacity wireless links, high-resolution imaging, and secure communications with spread spectrum or frequency hopping schemes.

However, several technical challenges remain in device development in this spectrum, which sets limits on power amplification techniques, detectors with large dynamic range, and efficient transceivers. Recent advancements in solid-state THz sources, graphene-based transistors, and photonic-integrated circuit architectures indicate significant progress toward overcoming these challenges. But it may take time for future WRCs and global markets to fully adopt these frequency bands for commercial wireless networks. 





\section{Future Research Directions for Emerging Spectrum Allocations and Coexistence}
\label{sec:openProblems}
The increasing demand for efficient spectrum utilization, coexistence with incumbent services, and regulatory adaptability necessitates innovative approaches in spectrum management, energy efficiency, and hybrid network architectures.
Given the existing and prospective frequency bands at UHF, mid-band, upper mid-band, and above 100 GHz for future wireless network design, several key research directions, including new scenarios of wireless communication system deployment in these emerging bands, intelligent spectrum allocation strategies that leverage artificial intelligence, consideration of energy efficiency in spectrum sharing, and integrated terrestrial and non-terrestrial networks, are expected to enhance spectrum sharing and adaptive allocation strategies. 


\subsection{New Scenario in the Upper Mid-band}
In addition to the commonly studied scenarios, such as urban microcell street canyon, urban macrocell, indoor office, and rural macrocell in 3GPP TR 38.901~\cite{38901}, an emerging scenario of suburban macrocell provide unique deployment opportunities in the upper mid-band spectrum, including high-speed broadband for residential areas, fixed wireless access as fiber alternative, and smart city applications. Existing channel models may fall short to faithfully characterize the channel statistics in this scenario with a sparse presence of residential structures, trees, and foliage, while an outdoor-to-indoor penetration loss study at 7--24 GHz under different materials and beamforming strategies can also be beneficial to comprehend the upper mid-band's potential in smart-home network applications.

\textcolor{black}{In a recently concluded channel modeling study (in June 2025) as part of 3GPP Release 19, researchers from Sharp, Nokia, Intel, and Ericsson, among other companies conducted field measurements and simulations at both sub-6 GHz and 7--24 GHz in suburban environments, to study the path loss, delay spread, LoS probability, and number of multipath clusters, among others~\cite{poddar2025overview}. A new plywood penetration loss model and an outdoor-to-indoor penetration loss model are also introduced in this study.} 

\subsection{AI-assisted Spectrum Allocation Strategies}
The integration of artificial intelligence (AI) and machine learning (ML) in spectrum management is a transformative step toward real-time, adaptive spectrum allocation. Recent advances in large language models (LLMs) and deep-learning algorithms can analyze historical spectrum usage data, learn spectrum access patterns, detect interference trends, and predict optimal frequency allocations, thus improving overall spectrum efficiency.

Moreover, deep reinforcement learning algorithms can facilitate environmental context-aware or semantic-aware spectrum decision-making, which is based on environmental conditions, network traffic demands, and regulatory constraints\textcolor{black}{~\cite{cao2024ai}}. However, a prerequisite is that the data required to train these AI-driven spectrum management systems must be carefully curated and validated to ensure accuracy, fairness, and compliance with regulatory standards. Open-source access to the data also allows fair comparisons and standardized assessment across different ML algorithms. 

\textcolor{black}{Future research needs to ensure that algorithms can recognize and address bias in AI models and provide interpretability~\cite{heydarishahreza2024spectrum}. Other critical challenges in the deployment of AI-assisted spectrum allocation strategies include protecting spectrum usage data privacy~\cite{perera2024survey}, adaptability of AI models in dynamic spectrum usage, and SDR system with intelligent spectrum agility~\cite{Madanayake2020towards}.}


\subsection{Energy-efficient Spectrum Sharing}

Energy efficiency is a crucial consideration in next-generation spectrum-sharing frameworks. The 6-GHz band (5.925--7.125 GHz) provides an example of licensed and unlicensed spectrum sharing by Ofcom in the U.K. and the FCC in the U.S.~\cite{ofcom2024,FCC_6GHz_2020}, where efficient spectral resource utilization is achieved by enforcing power and coordination constraints. In the 6-GHz band, unlicensed users operate in IEEE 802.11ax standard (also known as Wi-Fi 6), which provides a data rate at least twice as fast as that of the legacy Wi-Fi networks~\cite{FCC_6GHz_2020}, without causing interference to incumbent licensed users that operate wireless backhaul links at the 6-GHz band. This model serves as a blueprint for future spectrum-sharing strategies in higher-frequency bands, particularly in the upper mid-band, mmWave, and THz ranges.

Energy efficiency for NTN applications also remains an important issue. Unlike traditional terrestrial links, since the D2D links traverse the Earth's atmosphere, the operation of the non-terrestrial connectivity is subject to spectrum regulations and power spectral density limits to minimize the potential OOB interference with other space-borne services. 
Meanwhile, concerns about increased OOB interference not only arise from the other communication services (such as the ones in geostationary orbit), but also from the radio astronomy community~\cite{bassa2024bright}. The coexistence issue between terrestrial and non-terrestrial connectivity, as well as active communication and passive sensing services, will remain critical in 6G network deployments.

Additionally, future spectrum coexistence studies need to focus on optimizing transmission power according to real-time demand. The adoption of energy-efficient beamforming, AI-driven adaptive power control, and energy-efficient network architectures will be essential in reducing energy consumption while maintaining high spectral efficiency. The development of green spectrum-sharing policies by regulatory bodies (such as the recent FCC motion to review its rules on maximum transmissible power for satellites in the Ka- and Ku-bands~\cite{spacex2025powerloose}) around the world can also contribute to the achievement of sustainability goals in wireless communication networks. In a most recent proposed rulemaking by the FCC released on June 13, 2025 in modernizing spectrum sharing for satellite broadband~\cite{FCC062025SatelliteRule}, it is probable that the traditional equivalent power-flux density (EPFD) limits on geostationary and non-geostationary satellites that operate at 10.7--12.7 GHz, 17.3--18.6 GHz, and 19.7--20.2 GHz bands will be revised by the FCC and a degraded throughput methodology will be adopted. These frequency bands coexist with other services, such as radio astronomy and terrestrial applications. This recent FCC initiative will likely prompt broader ITU adoption of the degraded throughput criterion going forward.

\subsection{Integrated Terrestrial and Non-terrestrial Networks}
The convergence of terrestrial and non-terrestrial networks (NTNs) is another critical area of research to achieve seamless global connectivity in future wireless networks~\cite{geraci2022integrating, rahman2025joint}. Such an integration of satellites, high-altitude platforms (HAPs), and unmanned aerial vehicles (UAVs) with terrestrial infrastructure enables ubiquitous coverage, enhanced resilience, and extended service availability in remote and underserved areas.  Although there are existing partnerships and tests proven to be feasible in serving mobile users directly from satellites in low Earth orbits~\cite{tmobile2025}, data and broadband services are still not accomplished.

\textcolor{black}{In addition, several challenges in integrating terrestrial and non-terrestrial networks include vertical handover mechanisms across different networks for seamless mobility between terrestrial base stations and non-terrestrial nodes~\cite{warrier2024future, matera2024ground}, as well as latency and synchronization issues for applications requiring real-time communication~\cite{wang2025bridging, shang2024multi}, such as autonomous vehicles and emergency response networks.}

\subsection{\textcolor{black}{Coexistence of D2D with Ground-based Astronomy}}
\textcolor{black}{Previous studies have shown large deployable arrays designed for direct‑to‑device (D2D) links with noticeable optical brightness~\cite{nandakumar2023high}. For example, an observation campaign shows the peak magnitude of 0.4 for BlueWalker 3 satellite, making it one of the brightest objects in the sky~\cite{nandakumar2023high, Ravisetti2023BlueWalker3Brightest}. The  magnitude in astronomy refers to the Johnson-Cousins magnitude, which is a unit-less measure of the brightness of an object perceived in a defined spectrum~\cite{Bessell2005StandardSystems} and a lower value indicates higher brightness. The International Astronomical Union (IAU)'s Center for the Protection of the Dark and Quiet Sky from Satellite Constellation Interference (CPS) recommends that satellites in low Earth orbit (LEO) should not be visible by the unaided eye~\cite{boley2025iau}. More specifically, the CPS further recommends the magnitude of LEO satellites to be around 7.}

\textcolor{black}{The FCC Space Bureau has started the regulation process by including clear guidelines, including a recent authorization in August 2024 for AST to aim for 6th‑to‑7th‑magnitude brightness and to coordinate mitigation measures with the astronomy community~\cite{FCC_DA_24_756_AST_Brightness,FCC_DA_25_532_AST_Modification}. It is foreseeable that future launches will also abide by similar requirements to mitigate the brightness issues to ground-based astronomy.}

\section{Conclusion and Discussion}
\label{sec:conclusion}
This article outlines a vision of spectrum opportunities for the wireless future. Through surveying existing global allocations for passive and active services, interference considerations, and technical challenges related to propagation and hardware constraints, we identify potentially viable spectrum bands across UHF, upper mid-band, and above 100 GHz, as summarized in Table~\ref{tab:conclusion}. The findings provide valuable insights into the most promising frequency bands for future wireless systems, including direct-to-device satellite applications to 6G cellular services, while highlighting key regulatory and technological hurdles that must be addressed to enable the global adoption of upper mid-band and THz spectra for future wireless communications.

\begin{table}[h]
 \centering
    \caption{\textcolor{black}{A Summary of New Bands for 6G Wireless Networks and Beyond.}}
    \begin{tabularx}{\linewidth}{L|L|L|L|L}
    \hline
       \textbf{Frequency Range}  & \textbf{Possible Usage} & \textbf{Advantages} & \textbf{Drawbacks} & \textbf{Reference}\\\hline
       3.1--3.3 GHz & CBRS & US National Spectrum Strategy candidate & Partial overlapping with US DoD radar &~\cite{DoD2025Spectrum}\\\hline
      7.125--8.4 GHz (``golden band'') & Cellular &Favorable balance between propagation characteristics and coverage & Potential coexistence issue with 7.25--8.4 GHz satellite links & \cite{ntia2024plan}  \\\hline
      180 GHz (167--174.8 GHz) & Short-range ``whisper radio'' &  7.8 GHz available bandwidth & Limited propagation distance under wet air conditions & ~\cite{doc2021manual} \\\hline
      220 GHz (209--225 GHz) & Ultra-wideband fixed and mobile & 17 GHz contiguous bandwidth & Infrastructure and device readiness & ~\cite{li2024220} \\\hline
      280 GHz (275--296 GHz) & Ultra-wideband cellular & 21 GHz contiguous bandwidth, not allocated to specific services yet & Infrastructure and device readiness &~\cite{ETSI2024GR2} \\\hline
    \end{tabularx}
    \label{tab:conclusion}
\end{table}

As WRC-27 approaches, discussions on spectrum opportunities for the wireless future will continue to evolve. Consistent global regulations and cooperation will be essential to support the growing demand for next-generation wireless networks while protecting the needs of Earth observation and radio astronomy. Moving forward, the adoption of spectrum-sharing strategies, opportunities in the upper mid-bands,  the OOB interference challenge for fixed and mobile terrestrial links adjacent to satellite downlink, and D2D satellite service will be critical in maximizing the efficient use of radio frequency spectrum while addressing the concerns of both active and passive service stakeholders.

\section{Data Availability}
Data sharing is not applicable to this article as no datasets were generated or analyzed during the current study.

\section*{Acknowledgment}
This material is based upon work supported by the U.S. National Science Foundation under grants CNS-2216332, CNS-2339811, CNS-2403286, and CNS-2403287.

\section*{Author Contributions}
All authors, TSR, TEH, and SN, conceived the idea, performed the research, and wrote the manuscript. 

\section*{Competing Interests}
The authors declare no competing interests.

\bibliography{references.bib,pangea.bib}

\end{document}